\documentclass[smus]{snow2e}
\usepackage{epsf}

\begin{document}

\title{Discovery Mass Reach for Topgluons Decaying to $t\bar{t}$ at the Tevatron}

\author{Robert M. Harris\\ {\it Fermilab, Batavia, IL  60510}}

\maketitle

\thispagestyle{empty}\pagestyle{empty}

\begin{abstract} 
%
In topcolor assisted technicolor, topgluons
are massive gluons which couple mainly to top and bottom quarks.
We estimate the mass reach for topgluons decaying to $t\bar{t}$ at the
Tevatron as a function of integrated luminosity.
The mass reach for topgluons decreases with increasing topgluon width, and is 
$1.0 - 1.1$ TeV for Run II (2 fb$^{-1}$) and $1.3-1.4$ TeV for TeV33 (30 
fb$^{-1}$).

\end{abstract}
\section{Topcolor and Topgluons}
Topcolor assisted technicolor~\cite{ref_topcolor} is a model of dynamical
electroweak symmetry breaking in which the top quark is heavy because of a
new dynamics. Topcolor replaces the $SU(3)_C$ of QCD with $SU(3)_1$ for 
the third quark generation and $SU(3)_2$ for the first two generations. 
The additional SU(3) symmetry produces a $<t\bar{t}>$ condensate 
which makes the top quark heavy, and gives rise to a color octet gauge
boson, the topgluon B. The topgluon is expected to be wide 
($\Gamma/M \approx 0.3 - 0.7$) and 
massive ($M \sim 0.5 - 2$ TeV). In hadron collisions it is produced through 
a small coupling to the first two generations, and then decays via a much larger
coupling to the third generation: $q\bar{q} \rightarrow B \rightarrow 
b\bar{b}, t\bar{t}$. Here we estimate the mass reach for topgluons decaying to 
$t\bar{t}$ at the Tevatron.


\section{Signal and Background}

The sub-process cross section for $q\bar{q} \rightarrow t\bar{t}$ from both QCD
and topgluons  is given by~\cite{ref_lane}
\begin{equation}
\frac{d\hat{\sigma}}{d\hat{t}} = \frac{2\pi\alpha_s^2\beta_t}{9\hat{s}^2} (2
-\beta_t^2 +\beta_t^2\cos^2\theta^*)\left| 1 - \frac{\hat{s}}{\hat{s} - M^2 +
i\sqrt{\hat{s}}\Gamma} \right|^2
\label{eq_xsec}
\end{equation}
for a topgluon of mass $M$ and width $\Gamma$ given by 
\begin{equation}
\Gamma = \frac{\alpha_s M}{6} \left[ 4 \tan^2\theta + \cot^2\theta \left( 1 
+ \beta_t \left( 1-\frac{m_t^2}{M^2} \right) \right) \right]
\label{eq_width}
\end{equation}
where $\alpha_s$ is the strong coupling evaluated at renormalization scale
$\mu=m_t$, $\hat{s}$ and $\hat{t}$ are subprocess
Mandelstam variables, 
$\theta$ is the mixing angle between $SU(3)_1$ and $SU(3)_2$, 
$\theta^*$ is the scattering angle between the top quark and the initial state quark in the
center of mass frame, 
$\beta_t=\sqrt{1-4m_t^2/M^2}$, and $m_t$ is the top quark mass.
Topcolor requires $\cot^2\theta>>1$ to make the top quark heavy.
In Eq.~\ref{eq_width}, the first term in square brackets is for four light
quarks, and the second term has two components, the first for the bottom
quark and the second for massive top quarks.
In Eq.~\ref{eq_xsec}, the 1 inside the absolute
value brackets is for the normal QCD process $q\bar{q} \rightarrow g \rightarrow
t\bar{t}$.  The other term inside the brackets is the Breit-Wigner topgluon
resonance term for the process $q\bar{q} \rightarrow B \rightarrow t\bar{t}$.
The two processes interfere constructively to the left of the mass peak and
destructively to the right of the mass peak. 

In Fig.~\ref{fig_parton} we have convoluted Eq.~\ref{eq_xsec} with CTEQ2L
parton distributions~\cite{ref_cteq} to calculate 
the QCD background and topgluon signal
for the case of a 1000 GeV toplguon in $p\bar{p}$ collisions at $\sqrt{s}=2.0$ 
TeV. Fig.~\ref{fig_parton} also includes 
the QCD process $gg\rightarrow t\bar{t}$ which is only significant at low mass.
In Fig.~\ref{fig_parton}1a we plot the differential cross section $d\sigma/dm$,
where $m$ is the invariant mass of the $t\bar{t}$ system.
A clear distortion of the QCD $t\bar{t}$ 
spectrum is caused by the presence of a topgluon in Fig.~\ref{fig_parton}a. 
After subtraction of the QCD
background, Fig.~\ref{fig_parton}b shows that the signal has a very long high 
tail to low masses, 
caused by the combination of constructive interference and parton
distributions that rise rapidly as the $t\bar{t}$ mass decreases.  The
tail is larger than the peak, as seen in Fig.~\ref{fig_parton}b.
Nevertheless, the ratio between the topgluon signal and the QCD background,
displayed in Fig.~\ref{fig_parton}c, displays a noticeable peak close to the
topgluon mass. Similar calculations have been performed
for the masses 600, 800, 1200, and 1400 GeV.

\section{Discovery Mass Reach}

\begin{figure}[tbh]
\hspace*{-0.25in}
\epsffile{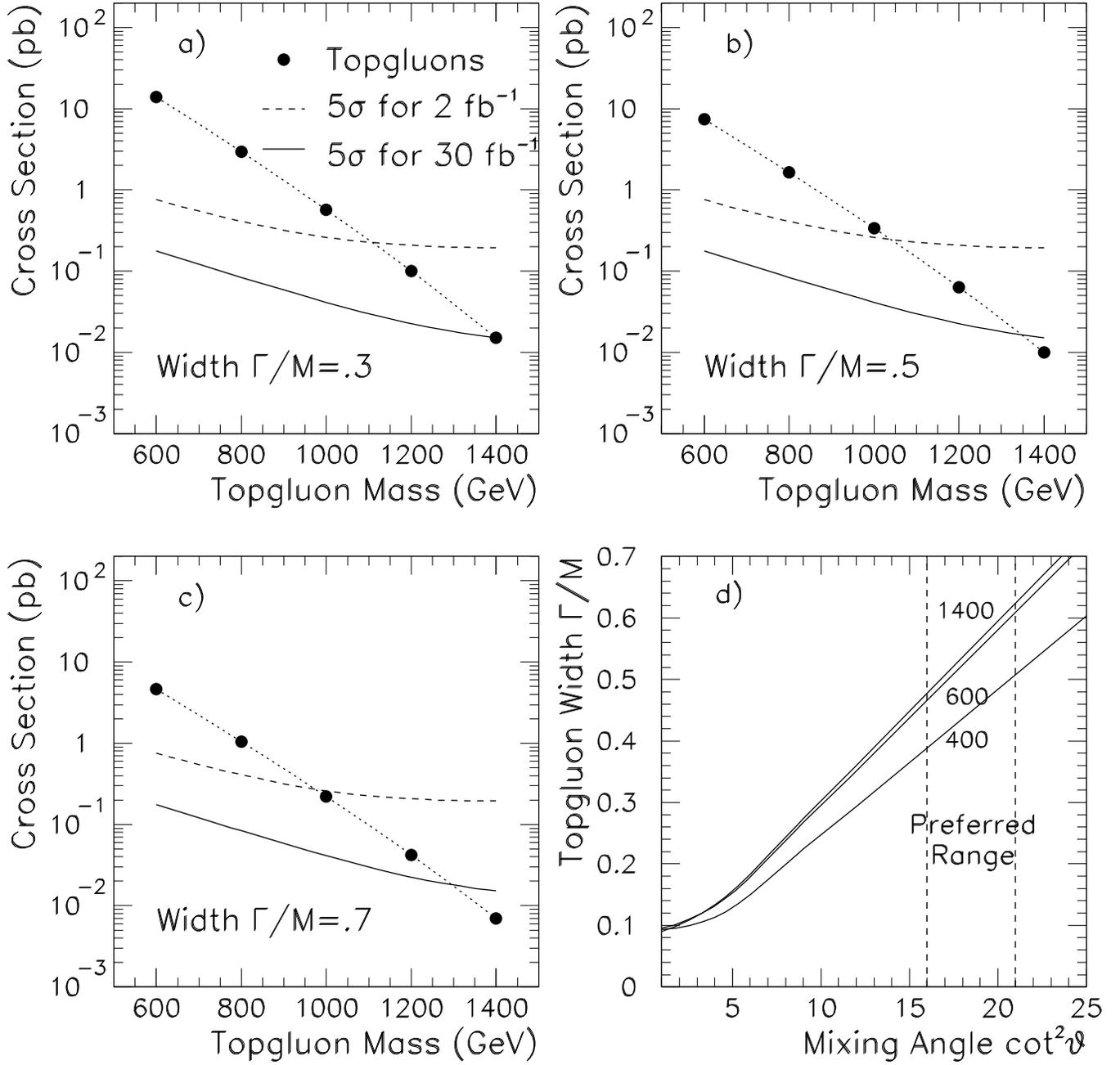}
\caption[]{ The mass reach for $b\bar{b}$ decays of topgluons of width a)
0.3 M, b) 0.5 M, and c) 0.7 M.  The predicted cross section for
topgluons (points) is compared to the 5$\sigma$ discovery reach of the Tevatron
with a luminosity of 2 fb$^{-1}$ (dashed) and 30 fb$^{-1}$ (solid). All 
cross sections are for $t\bar{t}$ with invariant mass within 25\% of the 
topgluon peak.  In d) the solid curves relate
topgluon width and the mixing angle, $\theta$, between $SU(3)_1$ and $SU(3)_2$ 
for 3 different topgluon masses. The vertical dashed lines indicate the 
theoretically preferred range of mixing angle~\cite{ref_topg_range}.}
\label{fig_reach}
\end{figure}

For topgluons of width $\Gamma/M\geq.3$, the measured $t\bar{t}$ mass peak should
have a resonance shape similar to the parton level distribution, since the 
detector mass resolution for $t\bar{t}$ ($\approx 6$\% at $m=800$
GeV) is significantly finer than the topgluon width.
To calculate the discovery mass reach we integrate both the lowest order 
topgluon cross section and the qcd background within the
range $0.75M<m<1.25M$.
The resulting total topgluon signal in the $t\bar{t}$ channel is shown
in Fig.~\ref{fig_reach}. The resulting background rate
in this mass range is used to find the 5 $\sigma$ discovery cross section. This
is conservatively defined as the cross section which is
above the background by 5 $\sigma$, where $\sigma$ is the statistical error on 
the measured cross section (not the background).  For example, if the 
background were zero events the $5\sigma$ discovery rate would be 25 events.  
To obtain the discovery cross section we used both the luminosity and
a 6.5\% $t\bar{t}$ reconstruction efficiency at CDF in run
II~\cite{ref_tev2000}.
In Fig.~\ref{fig_reach} we compare the
topgluon signal cross section to the 5 $\sigma$ discovery cross section for two 
different luminosities: 2 fb$^{-1}$ for Tevatron collider Run II and 
30 fb$^{-1}$ for TeV33. 
The topgluon discovery mass reach, defined as the mass 
at which a topgluon would be discovered with a 5$\sigma$ signal, 
is tabulated in Table I as a function of integrated luminosity and topgluon 
width.
The mass reach decreases with increasing width, caused by worsening 
\onecolumn
\begin{figure}[tbh]
\hspace*{-0.25in}
\epsffile{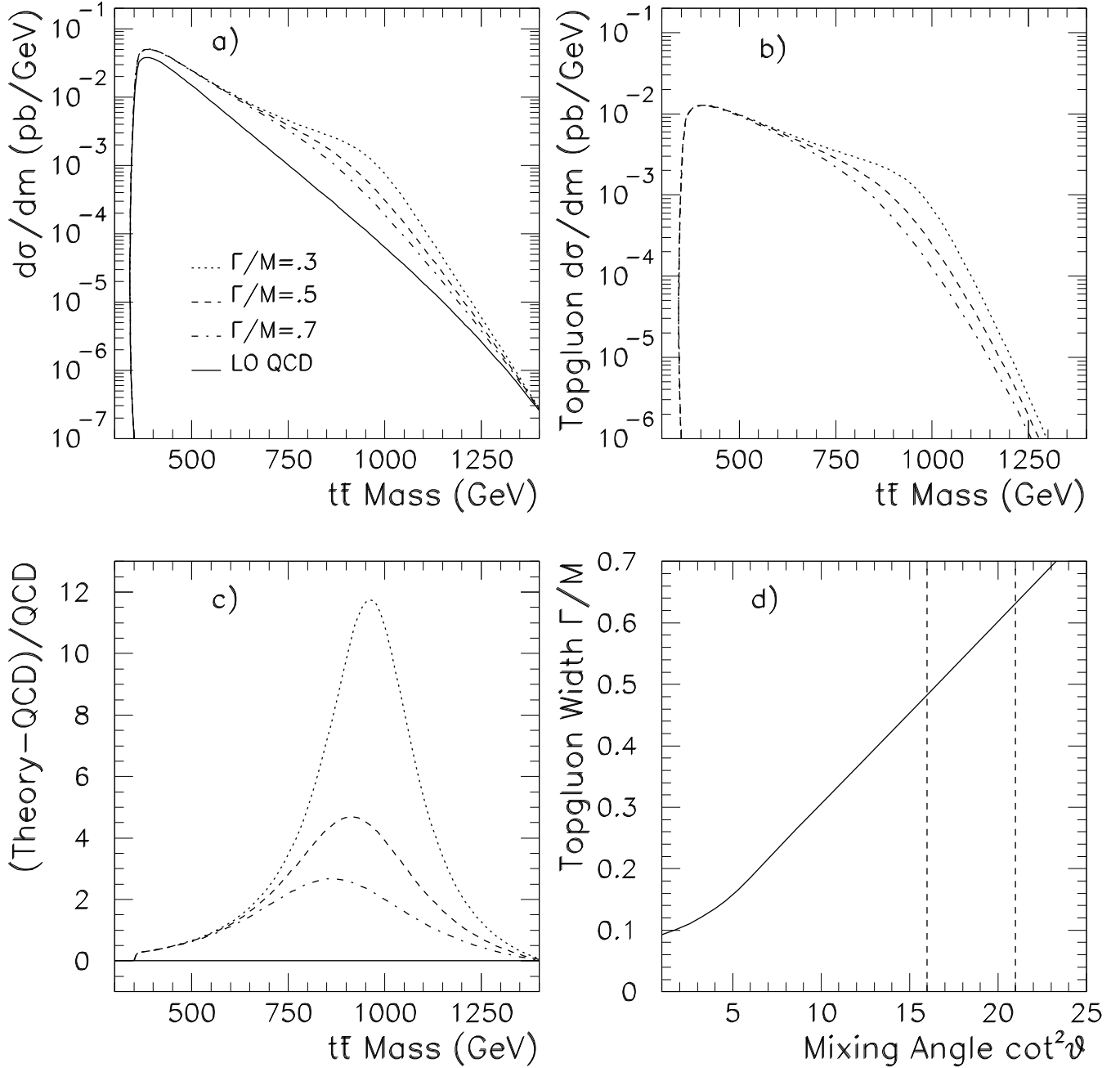}
\caption[]{ Lowest order parton level predictions for a 1000 GeV topgluon 
decaying to $t\bar{t}$ displayed as a function of $t\bar{t}$ mass. a) The 
cross section for the LO $t\bar{t}$ background 
from QCD (solid) is compared to the coherent sum of LO QCD and a topgluon
of fractional width $\Gamma/M=0.3$ (dots), 0.5 (dashes) and 0.7 (dotdash).
In b) the QCD prediction has been subtracted leaving only the topgluon signal
and the interference between QCD and topgluons (constructive beneath peak, 
destructive above peak). c) The fractional deviation above the QCD
prediction produced by the presence of a topgluon. d) The solid curve
relates the topgluon width and the mixing angle, $\theta$, between $SU(3)_1$ 
and $SU(3)_2$ for a 1000 GeV topgluon. The vertical dashed lines indicate the 
theoretically preferred range of mixing angle~\cite{ref_topg_range}.}
\label{fig_parton}
\end{figure}

\twocolumn
\noindent
signal to background within the search window. The width as a function
of mixing angle, from Eq.~\ref{eq_width}, is shown in Figs.~\ref{fig_parton}d 
and \ref{fig_reach}d. Also shown is the preferred theoretical range
for the mixing angle $\cot^2\theta$, determined from the topcolor model and 
constraints from other data~\cite{ref_topg_range}, which implies
a width of the topgluon in the range $\Gamma/M \approx 0.3 - 0.7$.  
\\
\\
\\
Table I: The 5$\sigma$ discovery mass reach of the Tevatron in Run II
(2 fb$^{-1}$) and TeV33 (30 fb$^{-1}$) for a toplguon decaying to $t\bar{t}$ 
as a function of its fractional width ($\Gamma/M$).
\begin{table}[h]
\begin{center}
\begin{tabular}{|c|c|c|}\hline 
Width & \multicolumn{2}{c|}{Mass Reach} \\
$\Gamma/M$ & 2 fb$^{-1}$ & 30 fb$^{-1}$ \\ \hline
0.3          & 1.11 TeV &  1.40 TeV \\
0.5          & 1.04 TeV &  1.35 TeV \\
0.7          & 0.97 TeV &  1.29 TeV \\ \hline
\end{tabular}
\end{center}
\end{table}           

\vspace*{-.1in}
\section{Systematics}
In the discovery mass reach estimate we have not included any systematic 
uncertainties on the measured signal, and we have assumed
that the shape and magnitude of the QCD $t\bar{t}$ spectrum will
be well understood.  We have not included any other sources of background,
such as QCD W + jets.  Also, our efficiency and resolution values are for 
reconstructing $t\bar{t}$ decaying into W + four jets where two of the jets
are b-tagged.  This efficiency and resolution may degrade at very high
$t\bar{t}$ mass values, because the byproducts from high $P_t$ top decay will 
be closer together, and modifications to the reconstruction technique may be 
necessary to preserve the efficiency and resolution. Adding systematics on the 
signal and the background will decrease the mass reach of a real search. 

\section{Total Cross Section Mass Reach}

\begin{figure}[tbh]
\epsfysize=3.5in
\epsffile{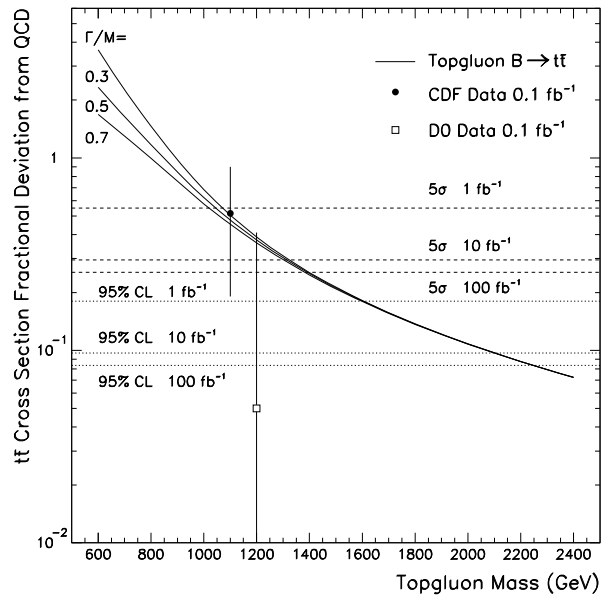}
\caption[]{ The fractional difference between the $t\bar{t}$ cross section
and the QCD prediction is shown for topgluons (solid
curves), CDF data (solid circle), and D0 data (open box).
The projected $5\sigma$ uncertainty (dashed lines) and 95\% CL (dotted lines)
on the measured $t\bar{t}$ cross section can be compared with the topgluon
prediction to determine the discovery reach and exclusion reach of the
Tevatron at the luminosities of 1, 10 and 100 fb$^{-1}$.}
\label{fig_total}
\end{figure}

Another method of searching for topgluons is simply to measure the total 
$t\bar{t}$ cross section and compare it with QCD. 
In Fig.~\ref{fig_total}
we show the fractional effect of a topgluon on the total $t\bar{t}$ cross 
section: (topgluon - QCD)/QCD. This is compared with the total $t\bar{t}$ 
cross section measurements
from CDF ($7.6$
\hspace*{-0.2cm} {\footnotesize $\begin{array}{c}{+1.9}\\{-1.6}\end{array}$}
pb~\cite{ref_cdf_xsec}).
and D0 ($5.2\pm 1.8$ pb~\cite{ref_tev2000}), both of which are compatible with QCD 
(5 pb~\cite{ref_xsec}) within errors.  In Fig.~\ref{fig_total} the location
of the CDF and D0 points on the horizontal axis is arbitrary; the measured
cross section and error give a location on the vertical axis only.
The TeV2000 group projected the $1\sigma$ uncertainty on the top quark cross
section measurement will be 11\% with 1 fb$^{-1}$, 5.9\% with 10 fb$^{-1}$,
and 5.1\% with 100 fb$^{-1}$~\cite{ref_tev2000}.  This estimate included 
systematic uncertainties.
In Fig~\ref{fig_total} we multiply these numbers
by a factor of 1.64 to obtain 95\% CL upper bounds, and multiply them by a
factor of 5 to obtain $5\sigma$ discovery cross sections, shown as horizontal dashed 
lines for luminosities of 1, 10 and 100 fb$^{-1}$. We interpolate between these
luminosity values to estimate the $5\sigma$ discovery cross section for 2 fb$^{-1}$
in Run II is 50\% of the QCD cross section, and the $5\sigma$ discovery cross 
section 
for 30 fb$^{-1}$ at TeV33 is 28\% of the QCD cross section.  These fractional
deviations in the QCD cross section correspond to a topgluon mass reach of 
$1.05-1.1$ TeV for 2 fb$^{-1}$, depending on the topgluon width, and about 
$1.35$ TeV for 30 fb$^{-1}$.  This assumes we understand the total QCD
cross section for $t\bar{t}$ production to much better than 50\% in Run II and
much better than 28\% at TeV33. This may not be unreasonable, considering the
theoretical systematic uncertainties on the $t\bar{t}$ cross section prediction
are currently around 10-20\%~\cite{ref_xsec}. Finally we note that the estimate
of the topgluon mass reach using the total $t\bar{t}$ cross section agrees 
with our estimate of the mass reach from the bump search in a 25\% mass window.
This increases our confidence that the estimated mass reach is reasonable.

\vspace*{-.1in}
\section{Summary and Conclusions} 
We have used a parton level prediction for $t\bar{t}$ production from QCD
and topgluons, together with the projected experimental efficiency for 
reconstructing $t\bar{t}$,
to estimate the topgluon discovery mass reach in a $t\bar{t}$ resonance search. 
The topgluon discovery mass reach,
$1.0-1.1$ TeV for Run II and $1.3-1.4$ TeV for TeV33, covers a significant part 
of the expected mass range ($\sim0.5 - 2$ TeV).  
The mass reach estimated using the total $t\bar{t}$ cross section is similar
to that for the resonance search, providing an important check.
For comparison, the mass reach in the $b\bar{b}$ channel is estimated to be
$0.77-0.95$ TeV for Run II and $1.0 - 1.2$ TeV for TeV33~\cite{ref_bbbar}.
This is less than the mass reach in the $t\bar{t}$ channel primarily 
because of larger $b\bar{b}$ backgrounds.
If topgluons exist, there is a good chance we will find
them at the Tevatron, beginning the investigation into the origins of 
electroweak symmetry breaking.

\end{document}